\documentclass[conference]{IEEEtran}
\IEEEoverridecommandlockouts

\usepackage{epsfig}
\usepackage{graphicx}
\usepackage[cmex10]{amsmath}
\usepackage[T1]{fontenc}
\usepackage{times}
\usepackage{ifthen}
\usepackage{latexsym}
\usepackage{afterpage}
\usepackage{verbatim}
\usepackage{cite}
\usepackage{url}

\fontfamily{ptm}\selectfont

\newboolean{draftmode}
\setboolean{draftmode}{false}

\begin{document}

\title{Iterative Detection and Decoding for the Four-Rectangular-Grain
TDMR Model \vspace{-0.15in}}

\author{\IEEEauthorblockN{Michael Carosino}
\IEEEauthorblockA{School of Electrical Engineering\\and Computer Science\\
Washington State University\\
Pullman, WA, 99164-2752\\
Email: mcarosin@eecs.wsu.edu }
\and
\IEEEauthorblockN{Yiming Chen}
\IEEEauthorblockA{Western Digital Corporation \\
Irvine, CA, 92612, USA \\
Email: yiming.chen@wdc.com }
\and
\IEEEauthorblockN{Benjamin J. Belzer, \\ Krishnamoorthy Sivakumar, \\ Jacob Murray and Paul Wettin}
\IEEEauthorblockA{School of Electrical Engineering\\and Computer Science\\
Washington State University\\
Pullman, WA, 99164-2752\\
Email: belzer,siva,jmurray,pwettin@eecs.wsu.edu }}

\maketitle

\begin{abstract}
This paper considers detection and error control coding for the 
two-dimensional magnetic
recording (TDMR) channel modeled by the two-dimensional (2D) 
four-rectangular-grain model
proposed by Kavcic, Huang et. al. in 2010. This simple 
model captures the effects of different 2D grain sizes
and shapes, as well as the 
TDMR grain overwrite effect: grains
large enough to be written by successive bits retain the polarity
of only the last bit written. We construct a 
row-by-row BCJR detection
algorithm that considers outputs from two rows at a time over
two adjacent columns, thereby enabling consideration of more 
grain and data states than previously proposed algorithms
that scan only one row at a time.
The proposed algorithm employs soft-decision 
feedback of grain states from previous rows to aid the estimation 
of current
data bits and grain states.
Simulation results using the same
average coded bit density and serially concatenated convolutional
code (SCCC) as a previous paper by Pan, Ryan, et. al.
show gains in user bits/grain of up to 6.7\% over the 
previous work
when no iteration is performed between the TDMR BCJR and the SCCC,
and gains of up to 13.4\% when the detector and the decoder iteratively
exchange soft information.     

\end{abstract}


\section{Introduction}
Industry is approaching the limit of the data storage density possible on
magnetic disk drives that write and read data on one-dimensional tracks.
Intensive efforts are underway in
alternative technologies such as heat-assisted-magnetic-recording
(HAMR) and bit patterned media recording (BPM). Most
of these techniques require
the recording medium to be radically redesigned \cite{Roger}, and
it is uncertain whether they will come on line quickly enough
to prevent a plateau in magnetic disk storage density in the near
to medium term.  

This paper considers detection and coding techniques
for an alternate approach proposed
in \cite{Roger} called 
two dimensional magnetic recording (TDMR), wherein bits are
read and written in two dimensions on conventional magnetic hard disks.
These disks have
magnetic grains of different sizes packed randomly onto the disk 
surface. In TDMR, information bits are channel 
coded to a density
of up to two bits per magnetic grain, and written by a special shingled
write process that enables high density recording.  A key problem
is that a given magnetic grain retains the polarization
of the last bit written on it; hence, if a grain is large enough to
contain two bit centers, the oldest bit will be overwritten by the newer
one.

A relatively simple model that captures the 2D nature of the TDMR
channel is the four-grain rectangular discrete-grain model (DGM) 
introduced
in \cite{Kavcic}, wherein four different grain types are constructed from
one, two, or four small square tiles.     
In \cite{Kavcic}, capacity upper and lower bounds for this 
model are derived 
showing
a potential density of 0.6 user bits per grain, translating to 
12 Terabits/$\mathrm{in^2}$ at typical media grain densities of 
20 Teragrains/$\mathrm{in^2}$. This is more than an order of magnitude improvement
over current hard disk drives, which exceed densities of
500 Gigabits/$\mathrm{in^2}$ \cite{LuPan-jour}.

\begin{figure}[b]
  \begin{center}
    \epsfig{figure=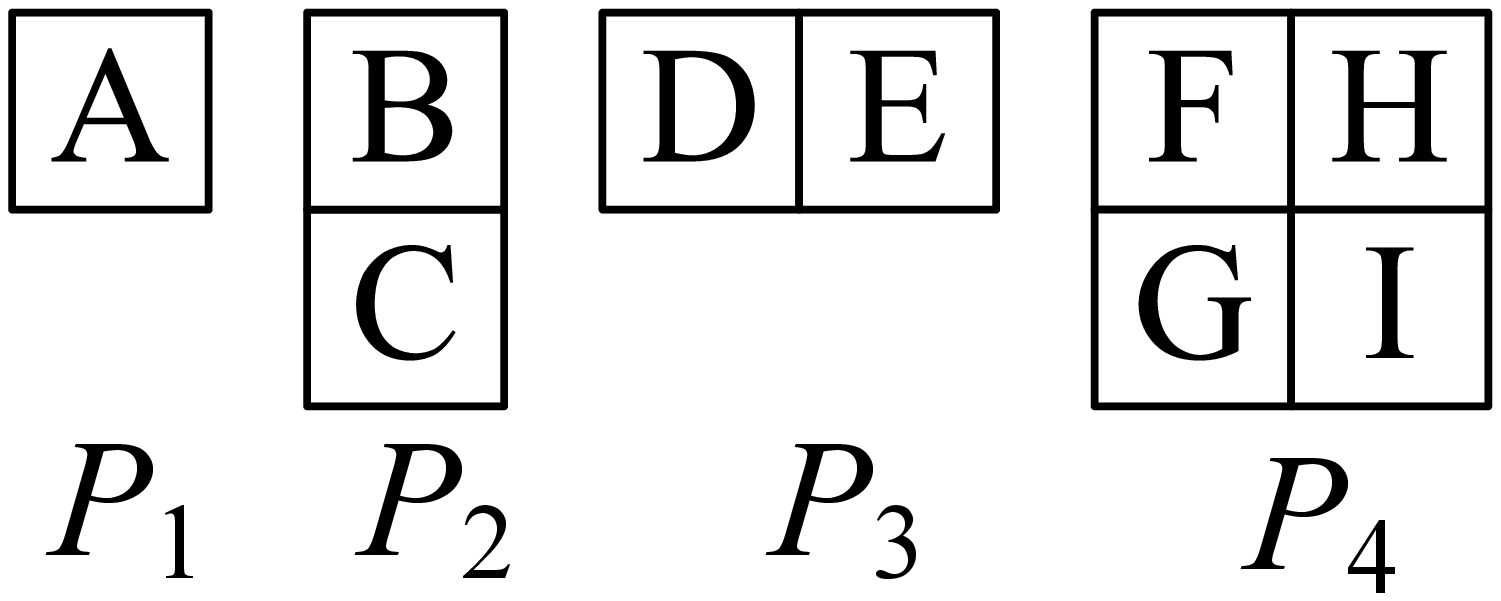,width = 2.0in}
    \caption{Four-grain rectangular discrete grain model assumed in this paper, from \cite{Kavcic}.}
    \label{fig: DGMmodel}
  \end{center}
\end{figure}

\begin{figure*}[t!]
  \begin{center}
    \epsfig{figure=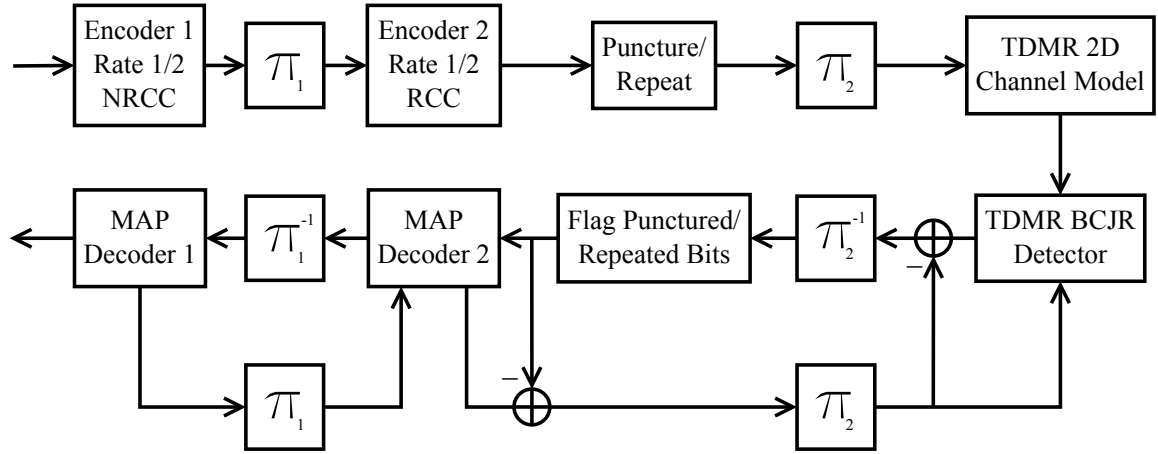,width = 6.0in}
    \caption{Block diagram of transmitter and receiver for SCCC coded
TDMR with iterative detection and decoding, after \cite{LuPan-jour}.}
    \label{fig: TDMR_SCCC_block}
  \end{center}
\end{figure*}

Coding and detection for the four-grain DGM is considered
in a previous paper by Pan, Ryan, et. al. \cite{LuPan-jour}. 
They construct a BCJR \cite{bcjr}
detection algorithm that scans
the input image one row (track) at a time. A 16-state trellis 
is constructed as the Cartesian
product of media states (that capture transitions between different
grain geometries during a one tile move along a row) and data states (that
capture the grain overwrite effect). It is shown that the number
of states can be reduced to 6 by combining equivalent states.
After one forward-backward pass through
each row of the input image, the TDMR detector passes soft information
in the form of log-likelihood ratios (LLRs) to a rate 1/4 serially concatenated convolutional
code (SCCC) with
puncturing, which decodes the data at the highest rate that achieves
a bit error rate (BER) of $10^{-5}$ (corresponding to the highest possible 
user bit density.)  No iteration between the TDMR detector and SCCC is done in
\cite{LuPan-jour}, although the possibility is mentioned.

This paper proposes a two-row BCJR detector for the
four-grain TDMR channel model. The novel contributions
are as follows:
1.) Considering the outputs of two rows and
two columns resulting from bits 
written on 
three input rows leads to a larger trellis
with more grain configurations, enabling
an increase in channel coding rates by up to 6.7\% over \cite{LuPan-jour};  
2.) Soft decision feedback of grain state information
from previously processed rows is used to aid the estimation of bits on
the current rows;
3.) States are labeled and enumerated to avoid geometrically
invalid states;
4.) By allowing the TDMR detector and SCCC decoder to 
iteratively exchange LLRs, code rate
increases of up to 13.4\% over \cite{LuPan-jour} are achieved.

This paper is organized as follows.  Section~\ref{sec: model_arch}
summarizes the four grain DGM and gives an overview of the proposed system
architecture.  Section~\ref{sec: bcjr_tdmr} explains the TDMR detection
BCJR algorithm. Section~\ref{sec: sims} describes simulation experiments,
and section~\ref{sec: conc} concludes the paper.

\section{Channel Model and System Architecture}
\label{sec: model_arch}

As shown in 
Fig.~\ref{fig: DGMmodel}, the rectangular DGM consists of four distinct grain
types consisting of unions of the smallest grain type; relative to the smallest
type their sizes are $1 \times 1$, $2 \times 1$, $1 \times 2$ and 
$2 \times 2$.  The four grain types occur with probabilities $P_1,\ldots,P_4$.
In this paper it is assumed that there is one channel coded bit per $1 \times 1$ grain,
and that the average number of coded bits per grain is 2, i.e., 
$1P_1 + 2P_2 + 2P_3 + 4P_4 = 2.$  Hence, if the channel coding rate is 
$r$ user bits per coded bit, the average number of user bits per grain 
is $2r$. The symmetry condition $P_2 = P_3$ is also
assumed; this, together with the two bits per grain condition, allows the
probabilities $P_1$ and $P_4$ to be computed given any valid value of $P_2 = P_3$.
The rectangular grains are packed at random according to their probabilities into a
$256 \times 512$ coded bit image, which models the magnetic disk surface, such
that every location $(m,n)$ in the image is covered by a grain or part of a grain. The subgrain labels A-F are used in the BCJR trellis definition described
in subsection~\ref{subsec: trellis}.

The system architecture employed in this paper is shown in Fig.~\ref{fig: TDMR_SCCC_block}.
A block of 32768 user information bits is 
encoded by a rate 1/4 SCCC from \cite{benser} consisting of an eight 
state rate 1/2 outer non-recursive
convolutional code (NRCC) with generator matrix $G_1(X) = [1 + X, 1 + X + X^3]$, followed by
an interleaver $\pi_1$, followed by an inner eight state recursive systematic convolutional code (RCC) with generator matrix $G_2(X) = [1, (1 + X + X^3) / (1 + X)]$, followed by a second interleaver $\pi_2$. Code rates greater than (respectively,
less than) 1/4 are achieved by randomly puncturing (respectively, repeating) 
randomly selected output bits from the inner encoder.
The code, input block size, puncturing/repeat scheme and output block
dimensions were chosen to be identical to those in \cite{LuPan-jour} in
order to facilitate comparison of the TDMR detector proposed in this paper with that
in \cite{LuPan-jour}.
The TDMR channel model writes coded bits taking the values $\pm1$ 
onto the $256 \times 512$ coded bit image row-by-row in raster scan order, with
multi-bit grains taking the sign of the last bit written on them. For example, all four
bits of the grain labeled ``FGHI'' in Fig.~\ref{fig: DGMmodel} become equal to
the bit written on its ``I'' subgrain, as this subgrain is farthest to the right and lowest and
hence is the the last written.

The lower half of Fig.~\ref{fig: TDMR_SCCC_block} depicts the 
iterative detection and decoding process. In the first outer
iteration of the entire system, the coded bit
image from the TDMR channel is read into the TDMR BCJR detector, 
which outputs channel bit LLRs to the inner MAP decoder.  The
inner MAP then exchanges LLRs with the outer MAP decoder for
several iterations before passing back estimates of the code
bit LLRs to the TDMR detector, and then another outer iteration
of the entire detector/decoder is started.  After several outer
iterations, the outer decoder puts out the decoded user bits.
Subtraction of incoming extrinsic information is performed
at the outputs of the TDMR detector and inner decoder before
they pass information to each other in order to
avoid feedback of previously processed outputs; similar
subtractions are performed in the SCCC decoder loop at the
outputs of the inner and outer MAP decoders, but these are not
shown in Fig.~\ref{fig: TDMR_SCCC_block}.

\section{BCJR Algorithm for TDMR Detection}
\label{sec: bcjr_tdmr}

\subsection{Trellis Construction}
\label{subsec: trellis}

\begin{figure}[h]
  \begin{center}
  \epsfig{figure=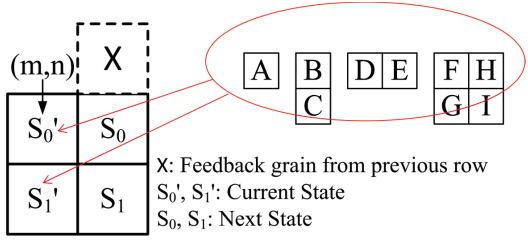,width=2.8in}
\caption{The grain state definition for the two-row TDMR
BCJR detector for the four-grain DGM of Fig.~\ref{fig: DGMmodel}.}
\label{fig: DGMreadback}
  \end{center}
\end{figure}

The grain state definition for the 
proposed BCJR detection algorithm is shown in Fig.~\ref{fig: DGMreadback}. 
Each of the state
subgrains can in theory take on one of the nine subgrain values A-I shown in 
Fig.~\ref{fig: DGMmodel}; however, grain connectivity restrictions shown in 
Table~\ref{table: ruleA} restrict the total number of current grain states
$(s_0',s_1')$ to 39. The bit ``X'' is feedback from the previously detected row;
the probabilities associated with the X bit are used to modify the state 
transition
probabilities in a soft-decision feedback scheme somewhat similar to the 
2D intersymbol interference (ISI) 
equalization algorithm described in \cite{taikun-spl}.

\begin{table}[h]
\caption{Connectivity restrictions between $s'_0$ at location (m,n) and 
$s'_1$ at (m+1,n), and between $s'_0$ at (m,n) and $s_0$ at (m,n+1).}
\label{table: ruleA}
\begin{center}
\begin{tabular}{|c|c|c|}
\hline 
$s'_0$: (m,n) & $s'_1$: (m+1,n) & $s_0$: (m,n+1) \\
\hline
A & A, B, D, E, F, H & A, B, C, D, F, G\\
\hline
B & C & A, B, C, D, F, G\\
\hline
C & A, B, D, E, F, H & A, B, C, D, F, G\\
\hline
D & A, B, D, E, F, H & E\\
\hline
E & A, B, D, E, F, H & A, B, C, D, F, G\\
\hline
F & G & H\\
\hline
G & A, B, D, E, F, H & I\\
\hline
H & I & A, B, C, D, F, G\\
\hline
I & A, B, D, E, F, H & A, B, C, D, F, G\\
\hline
\end{tabular}
\end{center}
\end{table}

\begin{figure}[h]
\begin{center}
  \epsfig{figure=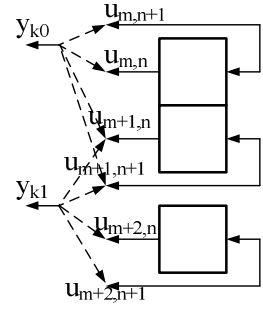,width=1.3in}
\end{center}
\caption{Data states for the 2-row BCJR detector.}
\label{fig: DGMdatastate}
\end{figure}

Fig.~\ref{fig: DGMdatastate} shows the data states for the 
two-row BCJR trellis.
This state-input
block scans through three input rows of a given 2D data block row-by-row in raster order, 
corresponding to the scan order of typical shingled writing heads proposed
for TDMR \cite{Voronoi_DGM_2}.  The trellis branch outputs 
$y_{k0}$ and $y_{k1}$ 
are the bits actually read from code bit locations $(m,n)$ and $(m+1,n)$, 
which are also the
location of the current state subgrains $(s_0', s_1')$.
The data states capture the
bit over-write property of TDMR, i.e., the relation between
the trellis branch outputs $(y_{k0}, y_{k1})$ and the corresponding 
input code bits that
were originally written on the disk. For example, output $y_{k0}$ may equal
input $u_{m,n}$ or ``future'' input $u_{m,n+1}$ depending on whether grain
states
$(s_0',s_0)$ are occupied by single
grains (such as AA, AB, etc.) or a connected grain (such as DE or FH).

Although trellis states can be constructed as the Cartesian
product of the media and data states, in fact there is only one possible
data state for each of the 39 media states shown in Table~\ref{table: ruleA},
so that the overall trellis needs only 39 states.  This is one of the
advantages of the subgrain labeling scheme in Fig.~\ref{fig: DGMmodel}:
it reduces the number of states in the initial trellis construction
compared to the scheme in \cite{LuPan-jour}, which constructs the trellis
as the Cartesian product of media and data states.  If we were to use
the subgrain labeling scheme to define the current and
next states $(s_0', s_0)$ employed in \cite{LuPan-jour}, we would require
only 9 states labeled A-I for $s_0'$, as opposed to the 16 states required
in the initial trellis construction in \cite{LuPan-jour}.  However, 
as pointed out in \cite{LuPan-jour}, combination of equivalent 
states in their trellis reduces the number of states from 16 to 6. Hence,
our subgrain labeling scheme does not give a minimum state trellis; 
however, it does provide a relatively straightforward method of
constructing trellises for multi-row TDMR BCJR detectors that are reasonably 
efficient in terms of state complexity and that are free of geometrically
invalid states.  

\subsection{Computation of BCJR Probabilities}
\label{subsec: bcjr_probs}

\renewcommand{\arraystretch}{1.4}
\begin{table*}[t]
\caption{Conditional probabilities of next states given that the current state is `AA'.}
\label{table: leastZero}
\begin{center}
\begin{tabular}{|c|c|c|c|c|c|c|c|}
\hline
$s$ & $P(s|s'=AA)$ & $s$ & $P(s|s'=AA)$ & $s$ & $P(s|s'=AA)$ & $s$ & $P(s|s'=AA)$\\
\hline
AA & $P_1\cdot P_1\cdot P(\bar{B},\bar{F})$ & AB & $P_1\cdot P_2 \cdot P(\bar{B},\bar{F})$ & AD & $P_1\cdot P_3\cdot P(\bar{B},\bar{F})$ & AE & 0\\
\hline
AF & $P_1\cdot P_4\cdot P(\bar{B},\bar{F})$ & AH & 0 & BC & $P_2\cdot P(\bar{B},\bar{F})$ & CA & $P_1\cdot P(B)$\\
\hline
CB & $P_2\cdot P(B)$ & CD & $P_3\cdot P(B)$ & CE & 0 & CF & $P_4\cdot P(B)$\\
\hline
CH & 0 & DA & $P_1\cdot P_3\cdot P(\bar{B},\bar{F})$ & DB & $P_2\cdot P_3 \cdot P(\bar{B},\bar{F})$ & DD & $P_3^2 \cdot P(\bar{B},\bar{F})$\\
\hline
DE & 0 & DF & $P_4\cdot P_3\cdot P(\bar{B},\bar{F})$ & DH & 0 & EA & 0\\
\hline
EB & 0 & ED & 0 & EE & 0 & EF & 0 \\
\hline 
EH & 0 & FG & $P_4\cdot P(\bar{B},\bar{F})$ & GA & $P_1\cdot P(F)$ & GB & $P_2\cdot P(F)$\\
\hline
GD & $P_3\cdot P(F)$ & GE & 0 & GF & $P_4\cdot P(F)$ & GH & 0\\
\hline
HI & 0 & IA & 0 & IB & 0 & ID & 0 \\
\hline
IE & 0 & IF & 0 & IH & 0 & & \\
\hline
\end{tabular}
\end{center}
\end{table*}

\begin{figure}[h]
\begin{center}
  \epsfig{figure=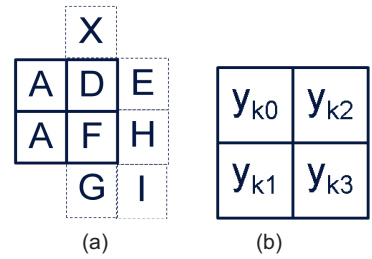,width=2.0in}
\end{center}
\caption{(a) Grain states involved in the transition
probability $P(DF | AA)$; note that the two A subgrains
are located at positions $(m,n)$ and $(m+1,n)$, corresponding
to the current state $(s_0',s_1')$ in Fig.~\ref{fig: DGMreadback}. (b) Outputs involved in
the conditional channel probability $P(\mathbf{y_{k}}\mid \mathbf{U=i},S_{k}=s,S_{k-1}=s')$; note that $y_{k0}$ and $y_{k1}$ are located at positions $(m,n)$ 
and $(m+1,n)$, respectively.}
\label{fig: TransProb}
\end{figure}

In the BCJR algorithm, the first and most important step is to compute 
the gamma state transition 
probability \cite{bcjr}:
\begin{equation}
\begin{split}
& \gamma_{\mathbf{i}}(\mathbf{y}_{k},s',s) = P(\mathbf{y}_{k}\mid \mathbf{U=i},S_{k}=s,S_{k-1}=s')\times \\
& P(\mathbf{U}\mid s,s')\times P(s\mid s').
\label{eq:gamma}
\end{split}
\end{equation}
Note that the {\em a priori} probabilities from the inner MAP decoder
have been left out of (\ref{eq:gamma}). We now describe the computation
of each of the factors in the gamma probability.
In (\ref{eq:gamma}), grain state transition probabilities $P(s\mid s')$ can be computed from 
Table~\ref{table: ruleA} and the grain probabilities shown in Fig.\ref{fig: DGMmodel}. 
The $P(s\mid s')$ probabilities can be stored in a $39 \times 39$ table.
Since the $P(s\mid s')$ table is very sparse, we 
show only one typical row in Table~\ref{table: leastZero}. In Table~\ref{table: leastZero}, P($\bar{B},\bar{F}$), 
P($\bar{B}$), P($\bar{F}$) specify probabilities of the feedback pixel 'X' in 
Fig.\ref{fig: DGMreadback}, 
where `$\bar{B}$' means `not equal to B'. These feedback probabilities
are computed from the LLRs from the detection of previous rows.  
Figure~\ref{fig: TransProb}(a) shows an example of the grain states
involved in the state transition probability $P(DF | AA)$, which is
calculated as:
\begin{equation}
P(DF | AA) = P_4 \cdot P_3 \cdot P(\bar{B},\bar{F}),
\end{equation}
where the last factor is the probability
that the
feedback subgrain X is neither a B nor an F
subgrain.

The computation of the feedback subgrain probabilities $P(X=B)$ and $P(X=F)$
proceeds as follows. 
We define the joint probability $\lambda$ in the usual way \cite{bcjr}:
\begin{equation}
\begin{split}
& \lambda^\mathbf{i}_k(s) = P(\mathbf{U}_k = \mathbf{i},S_k=s,\mathbf{y}^{N_r}_1) = \\
& \sum_{s'}\alpha_{k-1}(s') \cdot \gamma_{\mathbf{i}}(\mathbf{y}_{k},s',s) \cdot \beta_{k}(s),
\end{split}
\end{equation}
where we estimate two input bits $\mathbf{U}_k = (u_{0k},u_{1k})$ at 
each trellis stage, the $\alpha$ and $\beta$ probabilities are
defined as in \cite{bcjr}, and the notation $\mathbf{y}^{N_r}_1$ refers to
the entire sequence of received output vectors along one row of length $N_r$.
The LLRs $L(B)$ and $L(F)$ are computed as:
\begin{equation}
\begin{split}
& L(B) = \log \begin{bmatrix} \frac{\sum_\mathbf{i}\lambda_{k}^\mathbf{i}(s=BC)}{\sum_\mathbf{i}\lambda_{k}^\mathbf{i}(s\neq BC)} 
\end{bmatrix} \\
& L(F) = \log \begin{bmatrix} \frac{\sum_\mathbf{i}\lambda_{k}^\mathbf{i}(s=FG)}{\sum_\mathbf{i}\lambda_{k}^\mathbf{i}(s\neq FG)}
\end{bmatrix},
\end{split}
\end{equation}
where it is understood that the lambda probabilities are those for the 
feedback subgrain X. The probabilities $P(X=B)$ and $P(X=F)$ are then recovered
from the LLRs in the usual manner as
\begin{equation}
P(X=S) = \frac{\exp (L(S))}{1+\exp (L(S))},
\end{equation}
where $S$ is either $B$ or $F$.

The probability $P(\mathbf{U}\mid s,s')$ in (\ref{eq:gamma}) is equal
to 1/4, since the two coded bits in the input vector
$\mathbf{U}$ are assumed to be independent of each other (due to
the interleaver) and of the grain states.

The conditional probabilities 
$P(\mathbf{y}_{k}\mid \mathbf{U=i},S_{k}=s,S_{k-1}=s')$ of the $k$th output vector
$\mathbf{y}_{k} = (y_{k0},y_{k1},y_{k2},y_{k3})$ shown in 
Fig.~\ref{fig: TransProb}(b) are stored in a three dimensional array
of size $16 \times 4 \times 39 = 2496$, as the output vector
$\mathbf{y}_{k}$ depends only on the two element input vector and on the
previous state $S_{k-1}=s'$.  
In these probabilities, the input
bits $u_{k0}$ and $u_{k1}$ are assumed to be located 
at $(m,n)$ and $(m+1,n)$, i.e., input bit $u_{ki}$ is co-located with
output bit $y_{ki}$, for $i = 0,1$.
The conditional probabilities 
$P(\mathbf{y}_{k}\mid \mathbf{U=i},S_{k}=s,S_{k-1}=s')$
can take only the values of 0, 0.5, 0.25, or 0.125; in fact,
most of the probabilities in the table are equal to 0.  We now provide
examples of each of these four cases.  

First, 
$P(\mathbf{y}_{k}\mid \mathbf{U=i},S_{k-1}=FG)$ equals zero
whenever the four output bits in $\mathbf{y}_{k}$ are not all equal,
because in this case a single FGHI grain occupies the entire state block 
$(s',s)$ of Fig.~\ref{fig: DGMreadback}, and hence all the output bits
must be equal. This case can be detected easily because the algorithm
examines two rows and two columns at a time.
Second, for the same value of $S_{k-1}=FG$ as above,
$P(\mathbf{y}_{k}\mid \mathbf{U=i},S_{k-1}=FG) = 0.5$ whenever
the four output bits in $\mathbf{y}_{k}$ are all equal, because in
this case all four of these bits are determined by the single input 
bit at $(m+1,n+1)$, which is independent of either of the two given input bits
$u_{k0}$ or $u_{k1}$ at locations $(m,n)$ and $(m+1,n)$. This is the
only case in which the conditional probability takes the value 0.5.
Third, 
$P((u_{k0},u_{k1},y_{k2},y_{k3}) \mid (u_{k0},u_{k1}), S_{k-1}=AA) = 0.25$ for
any pairs $(u_{k0},u_{k1})$ and $(y_{k2},y_{k3})$, since in this case with 
$S_{k-1}=AA$ the input bits are written directly on the output bit locations, 
and the probability is equal to $0.25$ as long as the inputs and outputs agree.
(The presence of the two ``don't care'' bits $y_{k2}$ and $y_{k3}$, which can
assume four arbitrary values, accounts for the value of $0.25$.)  
Also, $P(\mathbf{y}_{k}\mid \mathbf{U=i},S_{k-1}=DD) = 0.25$ whenever
$y_{k0} = y_{k2}$ and $y_{k1} = y_{k3}$, and $0$ otherwise, since in this
case the two state rows are both DE grains, so that outputs $y_{k0}$ and 
$y_{k1}$ are determined by the two inputs
at $(m,n+1)$ and $(m+1,n+1)$ respectively, which can assume a total of
four different values. The only possible next state after DD is EE, so at
the next shift of the state register we will have the case
$P((u_{k0},u_{k1},y_{k2},y_{k3}) \mid (u_{k0},u_{k1}), S_{k-1}=EE) = 0.25$, similar
to the case above where $S_{k-1}=AA$. 
Fourth, 
$P(\mathbf{y}_{k}\mid \mathbf{U=i},S_{k-1}=DB) = 0.125$
whenever $y_{k0} = y_{k2}$, and 0 otherwise, 
since in this case there is a DE grain 
occupying the first state row $(s'_0,s_0)$ and
a BC grain occupying positions $(m+1,n)$ and $(m+2,n)$. Thus, outputs 
$y_{k0}$ and $y_{k1}$ are determined by the inputs at $(m,n+1)$ and 
$(m+2,n)$, which take four different values; the ``don't care'' bit
$y_{k3}$ can also assume two independent arbitrary values.

\section{Simulation Experiments}
\label{sec: sims}

In this section we describe
the details and give the results of Monte-Carlo simulations
of the system shown in Fig.~\ref{fig: TDMR_SCCC_block}.
The goal of the simulation experiments is to find the highest SCCC code rate that
allows decoding at a BER of $10^{-5}$ or lower, as higher code rates
correspond to a higher density of user bits (measured in user bits per magnetic
grain) on the magnetic disk. 

\subsection{TDMR/SCCC Interface and Iteration Schedule}
\label {subsec: det_dec}

\begin{figure}[h]
  \begin{center}
    \epsfig{figure=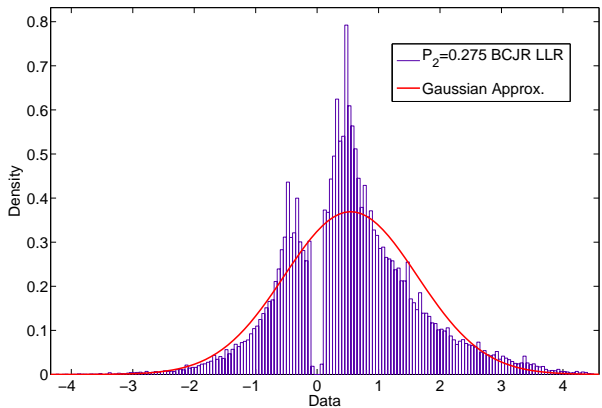,width = 3.7in}
    \caption{Experimental PDF of TDMR output LLRs conditioned on $+1$ 
bits, and best fit Gaussian PDF.}
    \label{fig: GaussianApprox}
  \end{center}
\end{figure}

With reference to the TDMR/SCCC detector/decoder shown in the bottom
half of Fig.~\ref{fig: TDMR_SCCC_block}, the LLRs output from the 
TDMR detector have a bimodal conditional PDF with two peaks
immediately to the right and left of zero, as shown in Fig.~\ref{fig: GaussianApprox}, 
which shows the experimental LLR PDF conditioned on correct bit values of $+1$. 
The LLRs around zero correspond to bits with low to
medium reliability; there is also a delta function at $+100$
(not shown in Fig.~\ref{fig: GaussianApprox}) that corresponds 
to high reliability bits estimated by the TDMR detector.
We find experimentally that modeling the central bimodal
PDF as a Gaussian as shown in Fig.~\ref{fig: GaussianApprox}, and using that Gaussian
as the conditional channel PDF in the SCCC decoder's BCJR algorithm, gives good results
for the SCCC decoder.
When no iterations are done between the TDMR detector and the SCCC decoder, we find
that optimizing the mean and the variance of the Gaussian model depending on the
value of probability $P_2$ (the probability of BC grains, and also of DE grains since
we assume $P_2 = P_3$) gives improved BER performance of the combined detector/decoder.
With iteration between the TDMR and SCCC, we find that we can fix the mean and
variance of the Gaussian model to be 1.0 and 1.69 respectively, for all values
of $P_2$.

With no TDMR/SCCC iterations, the SCCC decoder loop is run 30 times. This can
be reduced by using one of the several stopping criteria that have been developed
for turbo codes (e.g., \cite{shao}).  With iteration between the TDMR and SCCC, we
find experimentally that performing eight inner iterations of the SCCC decoder loop for each
outer iteration of the entire TDMR/SCCC loop gives the best overall BER performance.
The number of outer iterations required for successful decoding at a BER of $10^{-5}$
varies depending on value of $P_2$ (which also determines the code rate) and on 
the individual block being decoded, but 
never exceeds thirty outer iterations; the average number of outer iterations is
close to eight for all values of $P_2$ except $P_2 = 0$, where it is two. In place
of an SCCC stopping criteria, we check each decoded block for errors against the
known transmitted block, and stop decoding when zero errors are detected; although
this is not possible in an actual system, we believe that using a stopping criteria
would give similar results for the maximum and average numbers of iterations required.

\subsection{Monte Carlo Simulations}
\label {subsec: simres}

Each block of 32768 data bits is encoded into a block of $32768/r$ coded bits, where
$r$ is the code rate. For each block
of coded bits, a random grain image of size $256 \times 512 = 131072$ coded bits is
generated, corresponding to a code
rate of $r = 1/4$; higher or lower code rates are handled by deleting or adding rows
of length 512 from the random grain image. The random grain image is generated using
the greedy tiling algorithm described in \cite{Kavcic}, which places the largest 
grains first and then fills the holes between with smaller grains. 
However, at values of
$P_2$ greater than about 0.36, it becomes difficult to generate large random grain images
because there are not enough $1 \times 1$ grains to fill the holes between larger
grains. To overcome this, when $P_2$ is above 0.36 we generate a large set of $16 \times 16$
bit random sub-images, and then fill the large grain image with these $16 \times 16$ 
sub-images by randomly choosing sub-images from the generated set.  Once the random
grain image is generated, the coded bits are level shifted so that the values
$(0, 1)$ map to the values $(-1,1)$, and then the level shifted bits are
written onto the grain image in row-by-row raster
scan order, following the grain-overwrite rule that the last bit written onto a given
grain determines that grain's polarity.
The written grain image then flows into the TDMR/SCCC detector decoder.  A sufficient
number of data bit blocks are sent so that a BER of $10^{-5}$ or higher can be reliably 
measured.  In order to facilitate trellis termination, we surround each written grain image
by a boundary of $1 \times 1$ ``A'' grains that have all been written to $-1$.  To our
knowledge, these simulation conditions are exactly the same as those in 
\cite{LuPan-jour}, except that we are unsure how that paper handled grain image generation
at values of $P_2$ above 0.36.

\begin{figure*}[t]
  \begin{center}
    \epsfig{figure=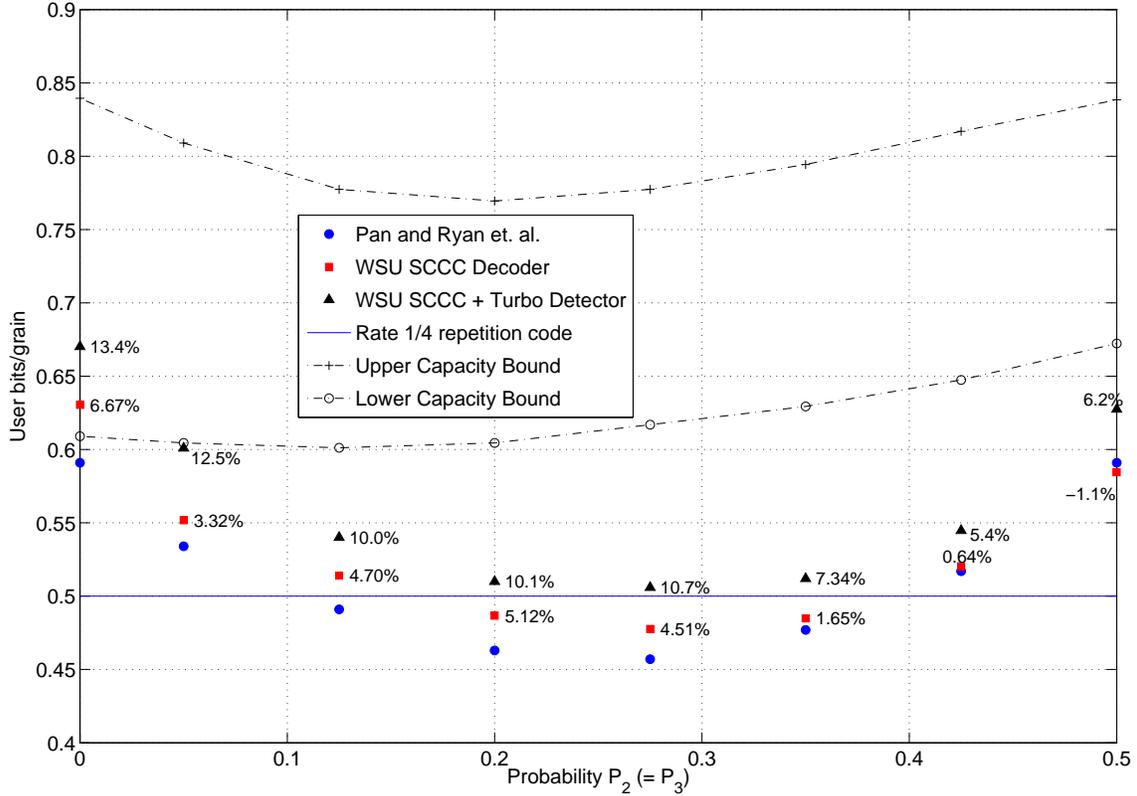,width = 7in}
    \caption{TDMR detection and decoding simulation results.}
    \label{fig: TDMRResults}
  \end{center}
\end{figure*}

The Monte-Carlo simulation results are shown in Fig.~\ref{fig: TDMRResults}. The figure's
horizontal axis is the probability $P_2$, and the vertical axis is the number of user
bits per grain, which is equal to twice the code rate since we assume a density of two
coded bits per grain.  The blue line at 0.5 user bits/grain corresponds to the bit
density that can be achieved by using a rate $1/4$ repetition code \cite{Kavcic}.
The upper and lower bounds on the channel capacity of the four-rectangular-grain 
TDMR channel (from \cite{Kavcic}) are also shown. 
The results achieved by the non-iterative TDMR/SCCC system of \cite{LuPan-jour} are
shown in blue dots; the corresponding results for the non-iterative system in the
present paper are shown in red squares, with the percentage rate gain relative
to \cite{LuPan-jour} shown immediately
adjacent to each red square. The largest rate gain of 6.67\% occurs at $P_2 = 0$, and
actually exceeds the lower capacity bound.  The rate gains decrease to around 5\% for $P_2$
between 0.2 and 0.3, and then decrease further until there is actually a rate loss of
about 1.1\% at $P_2 = 0.5$.  We believe the rate gains are largely due to processing two
rows (and four outputs) at a time in our BCJR algorithm; this allows, e.g., easy elimination 
of potential FGHI grain states by checking for agreement between all four bits in the grain. 
We are unsure about why the performance dips at $P_2 = 0.5$; possibly it is due to our using
a somewhat different method of generating random grain images at high values of $P_2$
than that used in \cite{LuPan-jour}.

The performance of our TDMR/SCCC system with iterative processing is shown in black 
triangles in Fig.~\ref{fig: TDMRResults}, with the rate gains relative to \cite{LuPan-jour}. 
The rate gains due to iterative processing are substantial. Again the best result is at $P_2 = 0$, where there is a 13.4\% rate gain; in fact, the point at $P_2 = 0$ attains 92.5\% of the average of the channel capacity 
upper and lower bounds, which is a rough approximation of the true 
channel capacity.  The rate gains in the most-difficult-to-handle middle range of 
$P_2$ between 0.2 and 0.3 are around 10\%; significantly, the achieved rates with iterative
processing are everywhere above the performance of the rate 1/4 repetition code.  
Also interesting is the performance at $P_2 = 0.5$, where a rate gain of 6.2\% is achieved;
iterative processing has raised the rate there by over 7.3\% compared to non-iterative 
processing.  The proposed iterative system achieves at least 72\% of the average of the 
channel capacity upper and lower bounds at all values of $P_2$; the corresponding figure 
for \cite{LuPan-jour} is 65\%. 

\section{Conclusion}
\label{sec: conc}

This paper has presented a system for detection and decoding of
bits on the four-rectangular-grain TDMR channel, using a novel
two-row BCJR-based TDMR detector, together with SCCC error correction
coding. Non-trivial gains in user bit density, especially at low values of
the two-grain probability $P_2$, are enabled by processing two
rows and four outputs at a time in the TDMR detector. 
When iteration is allowed between
the TDMR and SCCC, additional substantial density gains are enabled at
all values of $P_2$, albeit
at a complexity cost of roughly six or seven times that of the non-iterative
system, when the total numbers of SCCC and TDMR iterations are taken into
account.   

An interesting extension that we are currently pursuing
involves splitting the TDMR BCJR detector into two 
detectors, one that runs over rows as in this paper, and the other
that runs over columns. The two detectors can exchange soft bit 
and grain state information with each other and thereby possibly
converge at higher code rates then either could alone, as they
together can exploit the two different (and somewhat independent)
views of the TDMR channel. Similar row-column iteration strategies
have worked well in iterative equalizers for 2D intersymbol
interference channels 
(e.g., \cite{CC98,NXM00,WuOsull,marrowconf03,taikun-spl}). Results
of the row-column TDMR detector work will be reported in future publications.

\section*{Acknowledgment}

This work was supported by NSF grants CCF-1218885 and CCF-0635390.
The authors also wish to acknowledge useful discussions with 
Dr. Roger Wood of Hitachi Global Storage Technologies, San Jose, CA.

\bibliographystyle{IEEEtran}
\IEEEtriggeratref{9}
\bibliography{tdmr_ref,zzjour_chen_rev2_dir2disi-refer_corrected}

\end{document}